\documentclass[12pt]{iopart}

\usepackage{braket}
\usepackage{cite}
\usepackage{makecell}
\usepackage{color}
\usepackage{tikz}
\usepackage{cite}
\usepackage{iopams}
\usepackage{booktabs}
\usepackage{hyperref}
\hypersetup{
    colorlinks=true,
    linkcolor=blue,
    filecolor=blue,      
    urlcolor=blue,
    citecolor=blue
}
\begin{document}
	
	\title[Benchmarking quantum error-correcting codes on quasi-linear and central-spin processors]{Benchmarking quantum error-correcting codes on quasi-linear and central-spin processors}
	
	\author{Regina Finsterhoelzl and Guido Burkard}
	
	\address{Department of Physics, University of Konstanz, D-78457 Konstanz, Germany}
	\ead{regina.finsterhoelzl@uni-konstanz.de}
	
	\begin{abstract}
		We evaluate the performance of small error-correcting codes, which we tailor to hardware platforms of very different connectivity and coherence: on a superconducting processor based on transmon qubits and a spintronic quantum register consisting of a nitrogen-vacancy center in diamond. Taking the hardware-specific errors and connectivity into account, we investigate the dependence of the resulting logical error rate on the platform features such as the native gates, native connectivity, gate times, and coherence times. Using a standard error model parameterized for the given hardware, we simulate the performance and benchmark these predictions with experimental results when running the code on the superconducting quantum device. The results indicate that for small codes, the quasi-linear layout of the superconducting device is advantageous. Yet, for codes involving multi-qubit controlled operations, the central-spin connectivity of the color centers enables lower error rates.
	\end{abstract}
	
	\vspace{2pc}
	
	\section{Introduction}
	\label{sec:introduction}
	The observation and control of coherent quantum systems has advanced rapidly in recent years, leading to a quickened development of quantum technologies in the fields of quantum computing \cite{Ladd2010,Nielsen2010,Preskill1998}, quantum simulation \cite{Cirac2012,Bloch2012}, and quantum communication \cite{Kimble2008,Wehner2018}. Achievements in the area of quantum computing promise the possibility to ultimately perform computational tasks beyond the reach of high performance computers \cite{Arute2019}. To this end, physical platforms of very different properties are employed, ranging from photonic and atomic \cite{Bruzewicz2019,Monroe2021,Wu2021,Ebadi2021} to solid-state \cite{Awschalom2013,Burkard2021,Devoret2013,Kjaergaard2020,Rasmussen2021} systems.
	\begin{figure}
		\centering
		\includegraphics[width=0.7\linewidth]{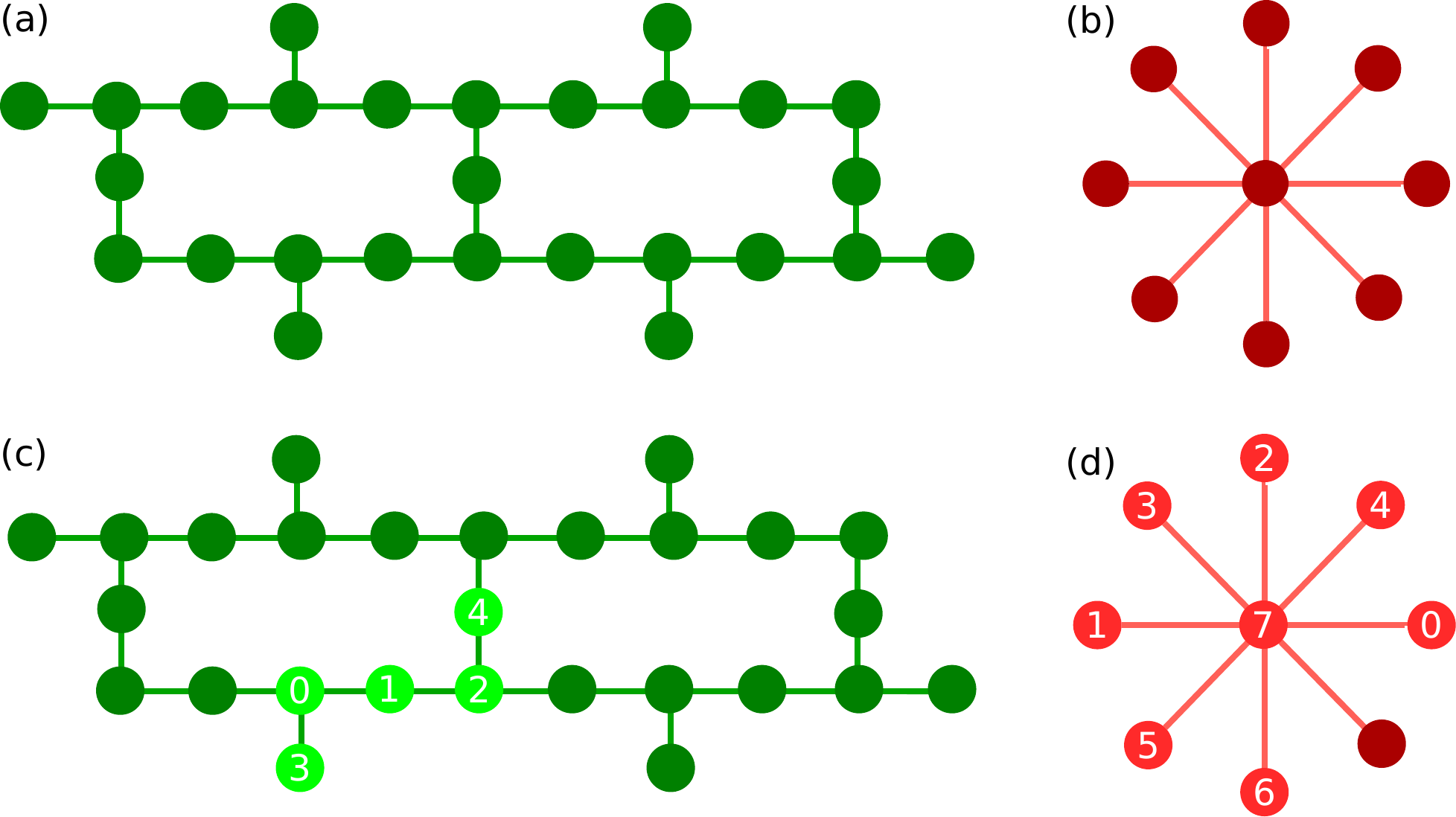}
		\caption{(a) Hexagonal-shaped coupling map of the IBM Q System One processor based on superconducting transmon qubits (SC-processor). (b) Central-spin coupling map of quantum register based on a nitrogen-vacancy center in diamond (NV-center). (c) Initial placement of the repetition code given in figure~\ref{fig:repetitioncode}a on the SC-processor, where 0,1,2 label the code qubits. A linear layout proved to be most advantageous for minimizing the number of swap operations during the execution of the code. (d) Initial placement of the same code on a quantum register built from a NV-center in diamond. Again, 0,1,2 label the code qubits. The electron spin in the center of the central-spin system (CSS) connectivity map is chosen as an ancilla qubit, while the role of the surrounding nuclear spins mainly depends on their respective state preparation and measurement errors and gate fidelities.}
		\label{fig:couplingmaps}
	\end{figure}
	However, these noisy intermediate-scale quantum (NISQ) devices are error-prone, making calculations on them imperfect due to gate infidelities and qubit decoherence \cite{Preskill2012, Preskill2018}. For a fully functional, fault-tolerant quantum computer, quantum error-correction (QEC) plays an essential part. It preserves coherence by spreading quantum information on physical qubits using entanglement \cite{Shor1996,Gottesman1997,Preskill1998,Nielsen2010,Lidar2013,Terhal2015,Campbell2017}. While the fault-tolerance threshold theorem \cite{Aharonov1997} proves that nearly noise-free computation using noisy components is possible with a moderate qubit overhead, the increase in code size nevertheless makes it difficult to implement QEC codes on NISQ hardware. This led to research efforts to reduce qubit overhead, for instance with the use of flag qubits or with topological codes in two or more dimensions \cite{Chao2018,Chao2020,Chamberland2018}. Recent research predicts requirements for fault-tolerant operations of such small codes or demonstrates their experimental implementation on various hardware platforms \cite{Abobeih2022,Chen2021,Krinner2021,Postler2022,Ryan-Anderson2021,Bermudez2019,Bermudez2017}.
	
	Theoretical predictions for the threshold of the physical error rate needed for fault-tolerant operations depend on the used code and error model and vary by several orders of magnitude \cite{Aliferis2008,Wootton2017,Wang2011}. However, in general, high-precision processor components with error rates below 1\% are required. Measuring their performance in a reproducible way is thus indispensable and is referred to as benchmarking \cite{Eisert2020,Gheorghiu2019}. Current state-of-the-art techniques include, e.g., randomized benchmarking \cite{Harper2019}, gate set tomography \cite{Blume2013} or quantum state tomography \cite{Rohling2022}, direct fidelity estimation \cite{Flammia2011} or cross-platform verification \cite{Elben2020}. These methods differ in complexity, assumption strength and information gain, and the obtained benefit depends on the problem at hand.
	
	The work presented in this paper investigates the performance of small error-correcting codes transpiled on two solid-state hardware platforms with very different native connectivity, gate sets, and coherence times: a processor built with superconducting transmon qubits featuring a quasi-linear connectivity and a hybrid spintronic quantum register based on defects in diamond with a central-spin system (CSS) connectivity (figure~\ref{fig:couplingmaps}). The target quantity for the comparison is the logical error rate, which we also use to benchmark our predictions with experimental results. We show that the superconducting processor achieves better error rates for small codes while for codes involving multi-qubit controlled operators with weight $w>2$, a fundamental operation not only in quantum error correction but also for instance in quantum simulation or fault-tolerant quantum computing \cite{Nielsen2010,Paetznick2013,Rasmussen2020}, the native gates and connectivity of the spintronic register proves advantageous. 
	
	Superconducting qubits belong to the technologically more advanced platforms \cite{Kjaergaard2020,Krantz2019,Devoret2004,Huang2020,Krastanov2015}. They consist of collective excitations in superconducting circuits, where the transmon qubit is built by a parallel circuit consisting of a nonlinear Josephson junction and a capacitor forming a non-equidistant energy spacing. The qubit levels are the lowest ones, resonant at $5\,\rm GHz$, where microwave pulses realize single-qubit operations \cite{Koch2007}. Transmon qubits feature a direct or mediated capacitive nearest-neighbour coupling \cite{Kjaergaard2020}, for instance realized by the cross-resonance gate, which is equivalent to a CNOT gate up to single qubit rotations \cite{Krantz2019, Rigetti2010,Chow2011,Barends2014}. This allows the construction of planar one- and two-dimensional arrays (figure~\ref{fig:couplingmaps}a). Readout may be performed using a linear superconducting resonator coupled to the transmon circuit. 
	
	The properties of defect-based quantum registers in solids differ in several aspects. Here, we consider the case of a register based on the nitrogen-vacancy (NV) center in diamond, where one carbon atom has been replaced by a nitrogen atom ($^{14} \rm N$) while one of its nearest-neighbour sites is vacant \cite{Pezzagna2021,Abobeih2022,Jelezko2006,Doherty2013,Dobrovitski2013}. Qubits are based on the NV electron spin as well as on the nuclear spins of the intrinsic $^{14} \rm N$ atom and the surrounding dilute $^{13}\rm C$ carbon atoms to which the electron spin couples via the hyperfine interaction. 
	A magnetic field lifts the degeneracy such that the qubit levels may be defined as $m_{s/N}\in [0,-1]$ with transition energies in the range of GHz (MHz) for the electron (nuclear) spins. Microwave (radio frequency) pulses are used for single- or multi-qubit conditional rotations \cite{Waldherr2014,Taminiau2014,Rong2015}. The individual sensing of up to 27 nuclear spins via the electron spin, the full control of up to 9 nuclear spins as well as ultra-long coherence times have been demonstrated experimentally \cite{Abobeih2022,Bradley2019,Abobeih2019,Zaiser2019,Waldherr2014}. The system is initialized \cite{Aslam2013,Weber2010} and read out \cite{Robledo2011,Abobeih2022,Abobeih2018,Vorobyov2013} optically via the electron spin and its coupling to the nuclear spins \cite{Cramer2016,Bradley2019,Abobeih2022}, enabling the native connectivity of a CSS where the electron spin mediates all couplings (figure~\ref{fig:couplingmaps}b). 
	Table~\ref{tab:platform_properties} compares fidelities and coherence times of both platforms.
	
	The experimental implementation of quantum error-correcting codes has been successfully demonstrated on both solid-state systems. While noise-resilient universal sets of single-qubit and multi-qubit gate operations \cite{Casanova2016,Shkolnikov2020} as well as its efficient use for quantum simulations \cite{Ruh2022} have been predicted for the NV-center, the minimal five-qubit code \cite{Bennett1996,Laflamme1996} has been implemented on a seven-qubit register \cite{Abobeih2022} making use of a recently proposed scheme using flag qubits. Prior to that, the three-qubit repetition code has been realized as a bit-flip or phase-flip correcting code\cite{Waldherr2014,Taminiau2014,Cramer2016}, and an encoding into a decoherence-protected subspace has been shown \cite{Reiserer2016}.
	On superconducting hardware, the three-qubit repetition code has been implemented successfully \cite{Reed2012,Riste2015}, while currently, the surface code is heavily explored, making use of the native 2d-planar connectivity of the transmon qubits \cite{Divincenzo2009,Corcoles2015,Gambetta2017,Chen2021}.
	
	This paper is structured as follows: in section~\ref{sec:error_model}, we introduce the two physical platforms used for the benchmark and explain the implemented error model. In section~\ref{sec:results}, we introduce the error-correcting codes and show our results, followed by a conclusion and outlook in section~\ref{sec:conclusion}.
	\begin{table}
	\caption{\label{tab:platform_properties}Comparison of important properties of two different hardware candidates for gate-based quantum computing: superconducting transmon qubits (SC) \cite{Koch2007,Barends2014,Kjaergaard2020} and spintronic qubits \cite{Abobeih2018,Bradley2019,Rong2015,Taminiau2014,Sar2012} consisting of a NV-center in diamond. $T_2^{\rm Hahn}$ denotes the spin-echo time measured by Hahn-type experiments, while $F_1$ ($F_2$) indicates the single-qubit (two-qubit) gate fidelity and $T_{\rm op}$ the operation temperature of the system. The spintronic hardware features very long coherence times (obtained at $3.7\,\rm K$). Contrary to the transmon processor which needs cryogenic technology and operation temperatures in the range of mK, the NV-center may be operated at higher temperatures up to $300\,\rm K$ (see table~\ref{tab:calibrationdata} for the corresponding parameters). While high-fidelity gates have been demonstrated both on the NV-center register as well as on transmon qubits, the former have been achieved at room temperature. The coherence time may be extended to over a minute for the nuclear spins \cite{Bradley2019} and to one second for the electron spin \cite{Abobeih2018} using dynamical-decoupling techniques. Note that the NV-center gate fidelities depend on the used spin type (see table~\ref{tab:calibrationdata}), while the fidelities listed here have been obtained for the electron and intrinsic nitrogen nuclear spin.}
	\centering
	\begin{indented}
        \item[]\begin{tabular}{@{}lllll}
		    \br
		    \lineup
		    qubit &&&&\\
		    \ns
			platform & $T_2^{\rm Hahn}$ (ms) & $F_1$ & $F_2$ & $T_{\rm op}$\\
			\mr
			\rule{0pt}{3ex}
			SC & 0.1 \cite{Kjaergaard2020, Barends2014} & 0.9992 \cite{Koch2007} & 0.994 \cite{Barends2014} & $15\,\rm mK$\\
			\mr
			\rule{0pt}{3ex}
			NV-center & \makecell[l]{Electron spin: 1.18 \cite{Taminiau2014}\\ $^{14}$N-spin: 2300 \cite{Abobeih2022,Bradley2019} \\
				$^{13}$C-spin: 260-770 \cite{Abobeih2022,Bradley2019} }& 0.99995 \cite{Rong2015} & 0.992 \cite{Rong2015} & $3.7$-$300\,\rm K$\\
		\br
		\end{tabular}
		\end{indented}
	\end{table}
	\section{Model and Methods}
	\subsection{Quantum Processors}
	\label{sec:quantum_processors}
	The prediction of the logical error rate achieved by the error-correcting codes is based on an error model built on calibration data. In order to benchmark this error model, we compare the simulation against the performance of the real quantum processor. For this, we make use of the latest processor generation built by IBM, the IBM Q Falcon processor \cite{Gambetta2022} which features a hexagonal connectivity \cite{Versluis2017}, see figure~\ref{fig:couplingmaps}a. In order to benchmark the performance of the transpiled codes, we also use the calibrated data of a specific NV-center operated by a group at the University of Stuttgart \cite{Zaiser2019}. Its native coupling map is CSS-like, where the central electron spins mediates all qubit-qubit interactions, see figure~\ref{fig:couplingmaps}b. The calibration data for both devices is listed in table~\ref{tab:calibrationdata}. Gate fidelities are obtained using randomized Clifford benchmarking techniques for both platforms \cite{Magesan2012,qiskit_benchmarking,Jaeger2021}.  
	
	Contrary to the SC-processor which is operated at $T_{\rm op}=15\,\rm mK$, the data for the NV-center is obtained at $T_{\rm op}=300\,\rm K$. State-preparation and measurement errors $p_{\rm spam}$ are in the range of a few percent for both platforms. Gate times $T_{\rm gate}$ are up to two orders of magnitude shorter for the SC-processor, while also the single (two)-qubit gate fidelity $F_1$ ($F_2$) differs in the same range, thus are more favourable on the SC-processor. Further improvement of the gate performance of the NV-center operations could be achieved e.g. using optimal control theory, as the higher fidelities listed in table~\ref{tab:platform_properties} indicate, which have been demonstrated at room temperature on the electron and nitrogen spin.
	Contrary to this, the coherence times are up to several orders of magnitude higher on the NV-center. The spin relaxation times $T_1$ depend the charge state and on the spin type of the NV-center. In the negatively charged (NV$^-$) state, the NV nuclear spins have relaxation times $T_1\gtrsim 250\,\rm ms$. 
	
	\begin{table}
	\caption{\label{tab:calibrationdata}Calibration data for the superconducting quantum processor IBM Q Falcon (SC-processor) \cite{Gambetta2022} and a quantum register based on a NV-center in diamond  \cite{Zaiser2019,calibrationdata_note}, both employed for benchmarking. $T_2^{\rm Hahn}$ times are spin-echo times obtained with Hahn-type experiments \cite{Zaiser2019,qiskit_t2}. Contrary to the SC-processor which is operated at $T_{\rm op}=15\,\rm mK$, the data for the NV-center is obtained at $T_{\rm op}=300\,\rm K$. State-preparation and measurement errors $p_{\rm spam}$ are in the range of a few percent for both platforms. Gate times $T_{\rm gate}$ are up to two orders of magnitude shorter for the SC-processor, while also the single (two)-qubit gate fidelity $F_1$ ($F_2$) differs in the same range and are thus more favourable on the SC-processor. The gate performance of the NV-center operations could be further improved e.g. using optimal control theory, as the fidelities listed in table~\ref{tab:platform_properties} - which have been demonstrated at room temperature - indicate. Contrary to this, the coherence times are up to several orders of magnitude higher for the NV-center. Note that $T_1$ strongly depends on the charge state and on the spin type of the NV-center.}
	\centering
		\begin{indented}
        \item[]\begin{tabular}{@{}llll}
		\br
		\lineup
    	&  \centre{2}{NV-center} & SC-processor\\
    	& electron spin & nuclear spin & \\
    	\mr
    	$T_1$ [ms] & 5.7 & $\gtrsim 250$ & 0.1 \\
    	$T_2^{\rm Hahn}$ [ms] & 0.4 & 0.9 & 0.1 \\
    	$T_{\rm gate}$ [ms] &  \centre{2}{0.001 - 0.15}  & 0.001\\
    	$1-F_1$ & \centre{2}{0.02} & 0.0008 \\ 
    	$1-F_2$ & \centre{2}{0.05 } & 0.01\\
    	$T_{\rm op}$  & \centre{2}{$300\,\rm{K}$}   & $15\,\rm{mK}$ \\
    	$p_{\rm spam}$ & \centre{2}{0.02-0.04} &0.02-0.04\\
    	\br
    \end{tabular}
		\end{indented}
	\end{table}

	Single native gates are the NOT gate (X) and its square root SX on the SC-processor, and rotations around the $x$-axis (RX) and $y$-axis (RY) on the NV-center. Native multi-qubit gates are the controlled-NOT gate (CNOT) on the superconducting hardware and the controlled-rotations gate (CROT) along the $x$- and $y$-axis on the NV-center. For both platforms, the RZ gate may be executed virtually by shifting the phase of the drive accordingly. As this does not require additional pulses, the gate is considered perfect ($F=1.0$) with zero gate time.
	
	\subsection{Error Model}
	\label{sec:error_model}
	To simulate the impact of decoherence on the circuit performance, we make use of an error model specified for each device building on calibrated data \cite{qiskit_github,Georgopoulos2021,Blank2020}. The errors are assumed to be uncorrelated and are described by noisy quantum channels $\mathcal{E}(\rho)$ acting on the density matrix $\rho$. In the operator-sum representation, they are given by
	\begin{equation}
		\mathcal{E}(\rho)  = \Tr_{\rm  env} \left[\mathcal{U} (\rho \otimes \rho_{\rm  env}) \mathcal{U}^\dagger\right] = \sum_k E_k\rho E_k^\dagger,
	\end{equation}
	where the Kraus operators $\{E_k\}$ fulfill $\sum_{k} E^\dagger_k E_k = \mathbb{I}$. Here, the map $\mathcal{E}(\rho)$ is completely positive and non-trace increasing. Note that the Kraus operators are not uniquely determined by $\mathcal{E}(\rho)$ \cite{Nielsen2010}.
	In addition to the SPAM errors, decoherence is modeled by taking qubit relaxation as well as errors due to faulty gates into account.
	Here, relaxation errors are assumed to occur due to amplitude damping and dephasing processes. Amplitude damping describes the effect of energy dissipation into the environment of a qubit \cite{Chirolli2008,Krantz2019}. Phase damping describes the loss of information about the relative phases between the energy eigenstates into an environment, but does not affect the population of the eigenstates. 
	When combined into a single quantum operation, the Kraus operators describing both amplitude and phase damping read as \cite{Ghosh2012} (see section \ref{sec:appendixkraus})
	\begin{eqnarray}
		E_{\rm apd,0}&=\left( \begin{array}{rr}
			1 & 0\\
			0 & \sqrt{1-p_{\rm ad}}\sqrt{1-p_{\rm pd}}
		\end{array}\right), 
		E_{\rm apd,1}=\left( \begin{array}{rr}
			0 & \sqrt{p_{\rm ad}}\\
			0 & 0
		\end{array}\right),\\
		E_{\rm apd,2}&=\left( \begin{array}{rr}
			0 & 0\\
			0 & \sqrt{1-p_{\rm ad}}\sqrt{p_{\rm pd}}
		\end{array}\right).
	\end{eqnarray}
	Here, we may relate the probability for the qubit to lose an excitation into the environment $p_{\rm ad}$ and the probability for the qubit to experience a random phase kick $p_{\rm pd}$ to the relaxation time $T_1$, the decoherence time $T_2$, and the gate time $\Delta t$, with 
	\begin{equation}
		\sqrt{1-p_{\rm ad}}= e^{\Delta t/2T_1}, \sqrt{1-p_{\rm ad}}\sqrt{1-p_{\rm pd}}= e^{\Delta t/2T_2}.
	\end{equation}
	Gate errors are captured with the depolarizing channel, where with probability $p_{\rm depol}$, the density matrix $\rho$ is replaced with a completely mixed state according to $\mathcal{E}_{\rm depol}(\rho) = p_{\rm depol}/d\,\mathbb{I} + (1-p_{\rm depol})\rho$. This results in the four Kraus operators $E_0 = \sqrt{1-3p_{\rm depol}/4}\mathbb{I},\,E_1=\sqrt{p_{\rm depol}}/2\,X, \,E_2=\sqrt{p_{\rm depol}}/2\,Y \, \textup{and} \,E_3=\sqrt{p_{\rm depol}}/2\,Z$,
	where $X,Y,Z$ represent the standard Pauli matrices $\sigma_x,\sigma_y,\sigma_z$, respectively. Lastly, SPAM-errors are modelled with a standard Pauli bit-flip channel which captures the probability of the qubit being prepared or measured in the $\ket{1}$ instead of the $\ket{0}$ state (and vice versa). It is applied before an ideal measurement operation $\mathcal{E}_{\rm m,ideal}$:
	\begin{equation}
		\mathcal{E}_{\rm m} = \mathcal{E}_{\rm m,ideal} \circ \mathcal{E}_{\rm spam}, \quad \mathcal{E}_{\rm spam}(\rho)=p_{\rm spam}X\rho X + (1-p_{\rm spam})\rho.
	\end{equation}
	Noisy single-qubit gates consist of an ideal, unitary gate $U$ followed by the quantum noise channels, thus $\mathcal{U} \rightarrow \mathcal{E}_{\rm damp} \circ \mathcal{E}_{\rm depol} \circ \mathcal{U}$. Errors on $N$-qubit gates $\mathcal{U}^{(N)}$ are assumed to occur uncorrelated on the single qubits (indexed with $i$) upon which $U$ acts non-trivially, thus
	\begin{equation}
		\mathcal{U}^{(N)} \rightarrow \bigotimes_{i=0}^{N-1} \mathcal{E}_{\rm damp,\textit{i}} \circ  \bigotimes_{i=0}^{N-1} \mathcal{E}_{\rm depol,\textit{i}} \circ \mathcal{U}^{(N)}.
	\end{equation}
	\subsection{Model parameterization}
    Based upon these assumptions, the model is adaptive to the respective hardware by capturing its native connectivity, coherence times and gate errors using calibrated data. To this end, we make use of the relationship between the average gate fidelity $F_{\rm av}$ and the Kraus operators of a quantum operation\cite{Watrous2018,Emerson2005} given by
	\begin{equation}
		F_{\rm av}(\mathcal{E}, U)
		= \int d\psi \langle\psi|U^\dagger
		\mathcal{E}(|\psi\rangle\!\langle\psi|)U|\psi\rangle = \frac{\sum_{k} |\Tr(E_k)|^2 + d}{d(d + 1)},
	\end{equation}
	where $U$ represents the target unitary of a noisy quantum channel $\mathcal{E}$ with dimension $d$. The average gate infidelity is given as the total gate error $p_{\rm g}$, which is obtained for each qubit and each native gate from calibrated data. We calculate the average gate fidelity due to relaxation processes $F_{\rm damp}$ from the parameters $T_1$, $T_2$ and $\Delta t$ and approximate the remaining gate error with $F(p_{\rm depol})=F[p_{\rm g}-(1-F_{\rm damp})]$. By also taking the native connectivity map into account, we obtain an error model specific to each device for the simulator.
	\subsection{Qubit routing}
	\label{sec:placement}
	The translation of a quantum circuit into a circuit adapted to the respective native gates, memory layout and error characteristics of a hardware platform is called transpilation or the qubit routing problem \cite{Svore2006,Siraichi2018,Cowtan2019,Leymann2020}. Transpiling a quantum circuit in an optimal way is crucial for the reduction of the impact of noise -- the task is to maximize the fidelity of the transpiled circuit. However, both finding the optimal initial layout as well as the optimal swapping sequences during the execution of the circuit are NP-hard combinatorial problems and thus come with very high computational costs, at least for larger circuits and devices. Additional resources are needed as both problems are intertwined and depend on the native gate set as well as on the gate fidelities of the given device. In recent years, many numerical solutions have been proposed for instance based on stochastic optimization \cite{qiskit_transpiler, Zulehner2019} or machine learning methods \cite{Note1, Pozzi2022}.
	
	In order to understand the hardware-specific impact of each of the limiting factors such as the native gate set, native topology and gate errors on the transpilation process, we transpile the circuit in steps, using analytical methods where possible, and check our result using numerical techniques provided by the open source framework qiskit \cite{qiskit_github}.
	First, we transpile the virtual circuit to the native gates, and minimize the circuit depth using circuit identities \cite{Nielsen2010} and the reduction of the number of circuit layers \cite{Svore2006}. We minimize the counts of the operation with the lowest fidelity, which are multi-qubit gates for both platforms. Where possible, we make use of mid-circuit projective measurements, combined with post-processing, to reduce the impact of noisy gates on the result.
	Next, we evaluate the optimal placement of the virtual qubits on the physical device. This corresponds to minimizing the number of additionally inserted SWAP gates while routing the qubits on the native topology of the device. As the SWAP operation is not native on either platform, it has to be transpiled at the cost of three CNOT gates with $\textup{SWAP}(q_0,q_1) = \textup{CNOT}(q_0,q_1)\textup{CNOT}(q_1,q_0)\textup{CNOT}(q_0,q_1)$ for two qubits labeled $q_0$ and $q_1$.
	
	We describe the native device connectivity as a directed graph, and for all subgraphs of different connectivity depending on the symmetry of the processor layout, we evaluate the initial layout requiring the least number of SWAP operations, which corresponds to providing the maximal number of native multi-qubit gates required in the specific circuit. When using post-processing, we choose the layout which enables us to perform as few operations after the projective measurement step as possible. 
	Finally, we use the error model to find the optimal placement of the subgraph on the hardware. Here, we calculate the average error rate $p_{\rm av}$ with $p_{\rm av} = 1-\Pi_{i=0}^{N-1} F_{i,\rm av}$ with $N$ the number of noisy operations in the transpiled circuit. Figure~\ref{fig:couplingmaps} depicts initial layout on the SC-processor (\ref{fig:couplingmaps}c) and on the NV-center (\ref{fig:couplingmaps}d) for an repetition code using five qubits.
	\section{Results}
	\label{sec:results}
	\subsection{Repetition Codes}
	\label{sec:bitflip}
\begin{figure}
		\centering
			\includegraphics[width=\textwidth]{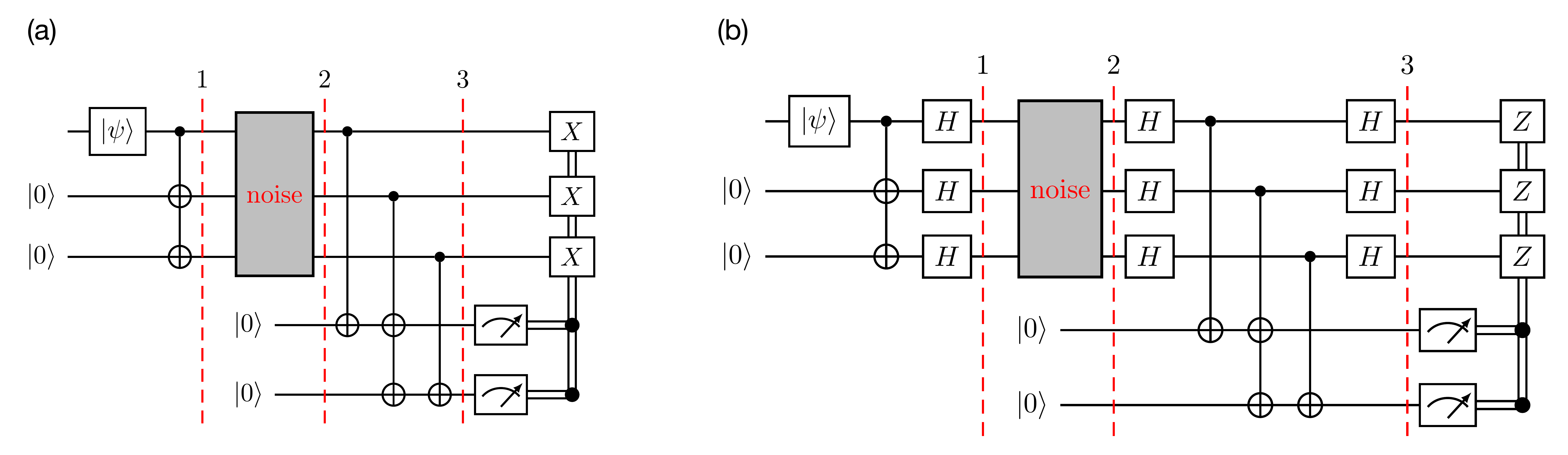}
		\caption{(a) The three-qubit repetition code correcting a single bit-flip error. It encodes the state $\ket{\psi}$ on a single physical qubit into the code space of three qubits with $\ket{\psi}=\alpha\ket{0}+\beta \ket{1} \rightarrow \ket{\psi}_L=\alpha\ket{000}+\beta \ket{111}\equiv\alpha\ket{0}_L+\beta \ket{1}_L$ (first part until line 1). Any bit-flip error occurring on the encoded qubit (line 1 - line 2) may be corrected by measuring the stabilizers $Z_1Z_2$ and $Z_2Z_3$ which will force the system into one of their eigenstates (until line 3). If $\ket{\psi}_L$ is in the code space, the measurement in the last section of the circuit will yield the eigenvalues $+1$ and leave the state undisturbed, while if an error has occurred, the measured syndrome will indicate the required recovery procedure. (b) The rotated three-qubit repetition code correcting single phase errors. Its structure is equivalent to the bit-flip code. First, the state $\ket{\psi}$ is encoded in the logical state space with $\ket{\psi}=\alpha\ket{0}+\beta \ket{1} \rightarrow \ket{\psi}_L=\alpha\ket{+++}+\beta \ket{---}\equiv\alpha\ket{0}_L+\beta \ket{1}_L$, and is then exposed to noisy evolution. With the following parity check using the stabilizers $X_1X_2$ and $X_2X_3$, phase-flip errors may be detected and corrected with Z-gates conditioned on the measurement result.}
		\label{fig:repetitioncode}
	\end{figure}

	We employ two fundamental error-correcting codes: the three-qubit repetition code fully correcting single bit-flip errors (bit-flip code) and the rotated three-qubit repetition code fully correcting single phase-flip errors (phase-flip code). Both codes have the advantage of requiring a low number of physical qubits. They are depicted in figure~\ref{fig:repetitioncode}a and figure~\ref{fig:repetitioncode}b.
	The repetition code encodes the state $\ket{\psi}$ on a single physical qubit into the code space of three qubits with $\ket{\psi}=\alpha\ket{0}+\beta \ket{1} \rightarrow \ket{\psi}_L=\alpha\ket{000}+\beta \ket{111}\equiv\alpha\ket{0}_L+\beta \ket{1}_L$, see figure~\ref{fig:repetitioncode}a. Any bit-flip error occurring on the encoded qubit may be corrected by measuring the stabilizers $Z_1Z_2$ and $Z_2Z_3$ which will force the system into one of their eigenstates. If $\ket{\psi}_L$ is in the code space, the result of both stabilizer measurements will yield $+1$ and leave the state undisturbed, while if an error has occurred, the measured syndrome will yield a negative eigenvalue $-1$ on one or both stabilizers. The result will indicate the required recovery procedure which implies a quantum operation conditioned on the measurement outcome.
	The phase-flip code may be described as a rotated bit-flip repetition code, where the quantum state is encoded as $\alpha \ket{0}+\beta\ket{1} \rightarrow \alpha \ket{+++} + \beta \ket{---} $. The stabilizers read as $X_1X_2, X_2X_3$ accordingly. The code is given in figure~\ref{fig:repetitioncode}b, its steps are equivalent to the ones of the bit-flip code explained above.
	
\begin{figure}
	\centering
		\includegraphics[width=\textwidth]{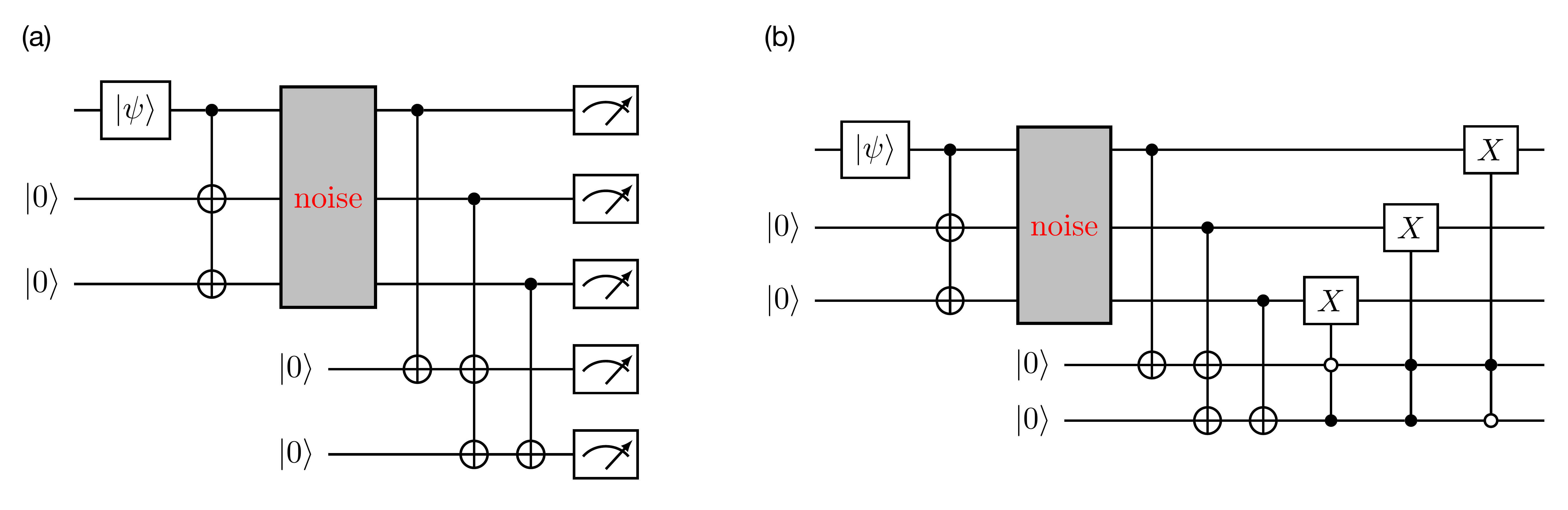}
		\caption{Error-recovery procedure without feed-forward on the example of the bit-flip code. (a) Classical post-processing, where the syndrome indicates a classical correction of the result of the quantum computation. (b) Circuit using unitary correction by multi-qubit conditioned quantum gates.}
		\label{fig:replace_feedforward}
	\end{figure}

	As it is generally the case in quantum error-correction, both circuits include a feed-forward operation, where the classical measurement result conditions the quantum operation during runtime. As this operation is not yet available on either of the processors employed here \cite{Note2}, we implement the recovery either by post-processing or by unitary correction. In the former case, the syndrome indicates a classical correction of the result of the quantum computation  (figure~\ref{fig:replace_feedforward}a), while the latter implements the correction using multi-qubit conditioned quantum gates (figure~\ref{fig:replace_feedforward}b).
	
	\subsection{Benchmarking the model}
	We use the bit-flip code including post-processing for benchmarking the error model. To this end, we deliberately induce a random bit-flip error on one of the code qubits during the noise evolution and evaluate its correct detection using the syndrome.
	We transpile the circuit to both hardware platforms (see section~\ref{sec:placement}) using the calibrated data described in section~\ref{sec:quantum_processors}.
	\begin{figure}
		\centering
		\includegraphics[width=0.9\linewidth]{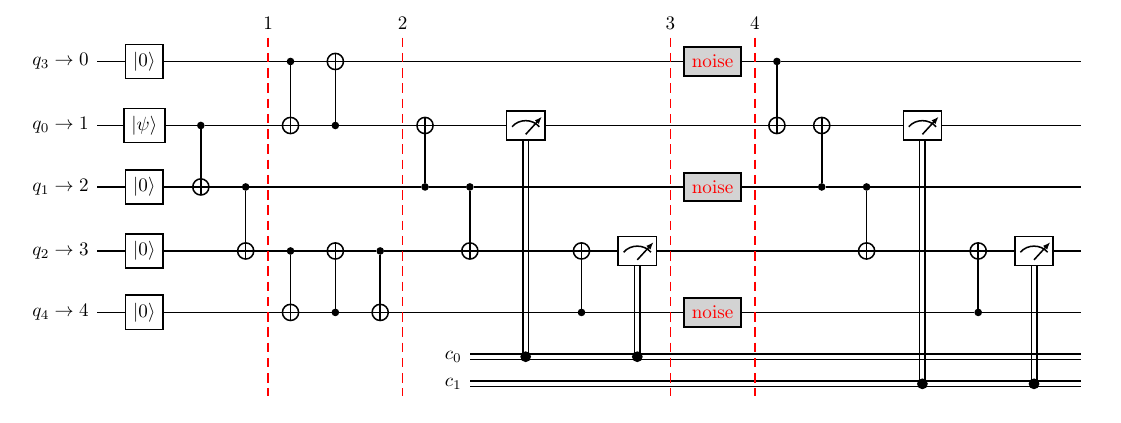}
		\caption{Transpiled bit-flip code given in figure~\ref{fig:repetitioncode}a on the superconducting processor using the initial layout depicted in figure~\ref{fig:couplingmaps}c. The number of CNOT gates in the code is minimal when transpiled to the linear connectivity, as this layout provides all native two-qubit gates which are required after the intermediate measurement step.}
		\label{fig:BitFlipIBM}
	\end{figure}
	On the given topology of the superconducting device, all connected subgraphs are of linear or nearly linear connectivity. Interestingly, the number of CNOT gates in the code is minimal when transpiled to the linear connectivity, as this layout provides all native two-qubit gates which are required after the intermediate measurement step. Also, it enables multi-qubit gate cancellation, as two sequential identical CNOTs form the identity according to $\rm CNOT(q_0,q_1)\rm CNOT(q_0,q_1)=\mathbb{I}$. 
	Figure~\ref{fig:couplingmaps}c depicts a possible initial placement of the virtual qubits on the physical device. The resulting circuit transpiled for the superconducting device is depicted in figure~\ref{fig:BitFlipIBM}.
	
\begin{figure}
		\centering
		\includegraphics[width=0.9\linewidth]{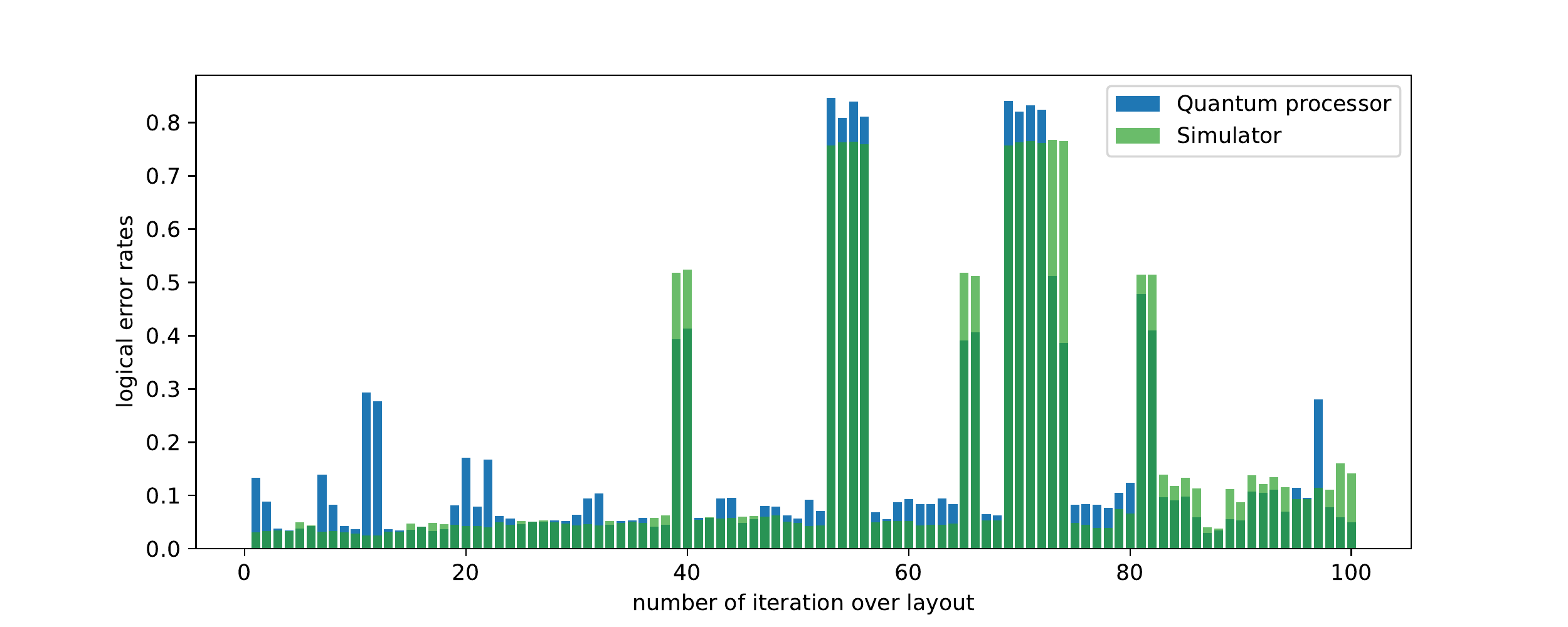}
		\caption{Benchmark of the simulation using the error model described in section~\ref{sec:error_model} with the performance of the code on a real superconducting quantum device. The model slightly underestimates the error, but captures the behaviour of the noisy processor in most cases. Deviations are presumably due to hardware-specific, correlated errors.}
		\label{fig:comparison_sim_exp}
	\end{figure}

	We benchmark the noise model with the behaviour of the real quantum device. The results are depicted in figure~\ref{fig:comparison_sim_exp}. The model qualitatively captures the behaviour of the quantum device. Here, the error rate predicted by the simulator generally lies slightly below the one obtained by the experiment, while for some cases, the prediction is too low. This is because the model is built from Markovian error types which are hardware agnostic and thus fail to capture hardware-specific, spatially correlated processes such as cross-talk \cite{Tripathi2021}. It could be adapted to the specific hardware for instance using neural networks trained to the specific errors as demonstrated in \cite{Georgopoulos2021}. However, when using the error model for benchmarking we require it to remain on the more general level given here.
	
	\subsection{Performance of the bit-flip code}
	\label{sec:bitflipperformance}
\begin{figure}
		\centering
		\includegraphics[width=0.9\linewidth]{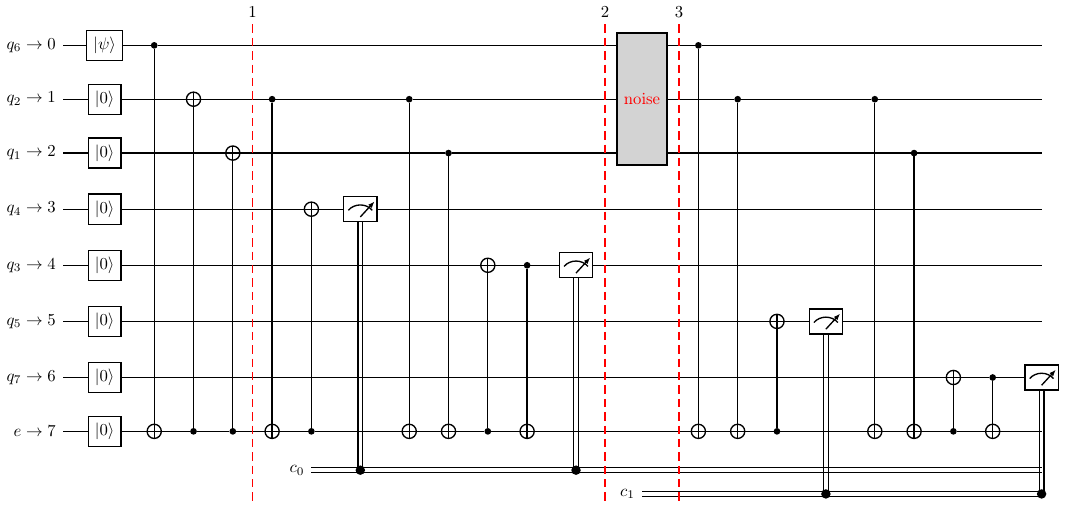}
		\caption{Transpiled bit-flip code given in figure~\ref{fig:repetitioncode}a on a quantum register built by a NV-center in diamond using the initial layout depicted in figure~\ref{fig:couplingmaps}d.}
		\label{fig:BitFlipNVCenter}
	\end{figure}

\begin{figure}
		\centering
		\includegraphics[width=0.7\linewidth]{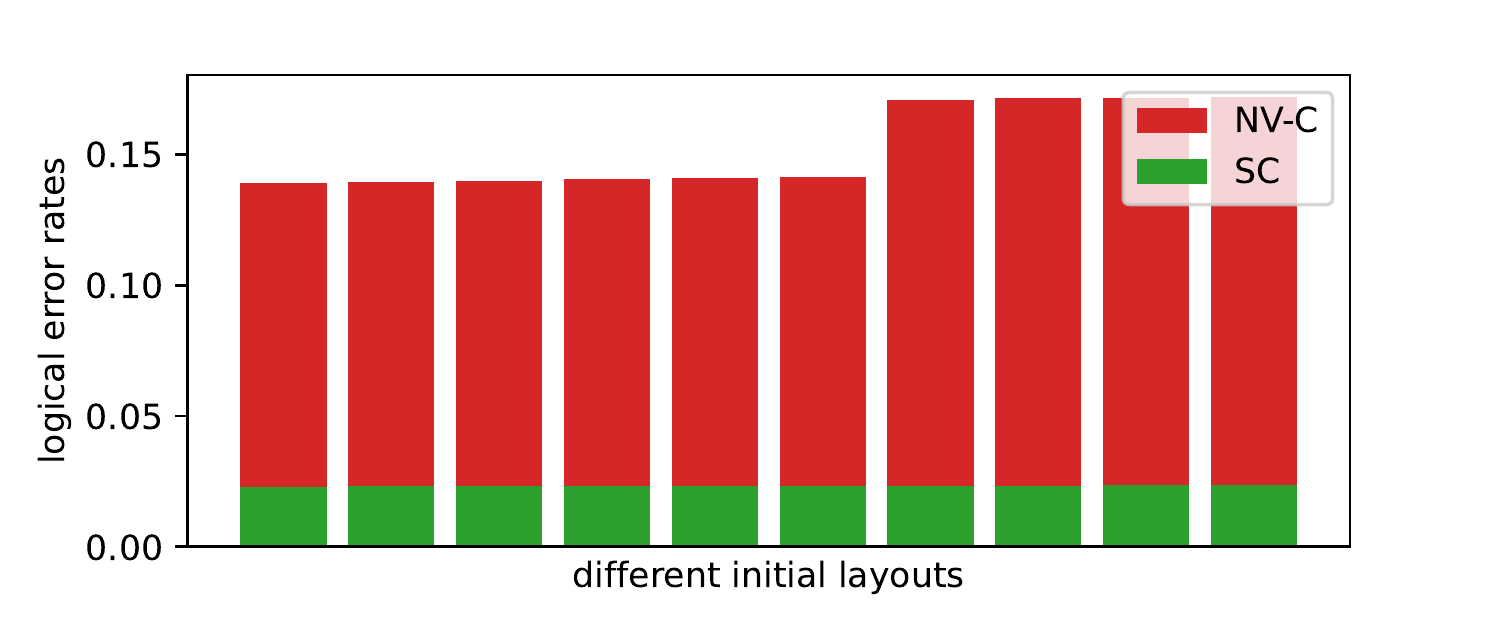}
		\caption{Comparison of best logical error rates achieved by the bit-flip code on a superconducting processor (SC) and a NV-center quantum register (NV-C). Initial state is $\ket{0}$. Results for other pure initial states agree up to a deviation of around 3\%.}
		\label{fig:tenbestresults}
	\end{figure}

	Next, we use the benchmarked error model to evaluate the performance of the bit-flip code comparatively on both platforms. To this end, we transpile the code also to the NV-center native properties.
	Figure~\ref{fig:BitFlipNVCenter} depicts the transpiled circuit, while figure~\ref{fig:couplingmaps}d depicts the initial layout. The central electron spin is chosen as ancilla qubit. Again, the first part until line 1 depicts the encoding, the projection into code space (line 1-line 2) followed by the random bit-flip error (line 2-line 3). The part until line 4 depicts the stabilizer measurement. Due to the native connectivity, the central spin serves as a mediator both for the entangling gates and readout. Note that the CSS-like connectivity map is less favourable in this case when compared to the hexagonal layout the SC-processor. While in the latter case, the transpiled circuit only requires three CNOTs for measuring the stabilizers, their number rises to seven CNOTs in the case of the NV-center register.
	Again, we run the code on the simulator. The best results of both systems are depicted in figure~\ref{fig:tenbestresults} for two different initial states. The code performs better on the superconducting hardware. Here, the best results reach a logical error rate of 0.023 compared to 0.139 on the NV-center.
	In order to compare the influence of certain types of errors - those due to limited coherence times and those due to gate imperfections - we evaluate the impact of both errors separately. To this end, we build the model with equal parameters for both systems, enabling a comparison of the influence of the native gates and native connectivity.
	\begin{figure}
		\centering
		\includegraphics[width=\textwidth]{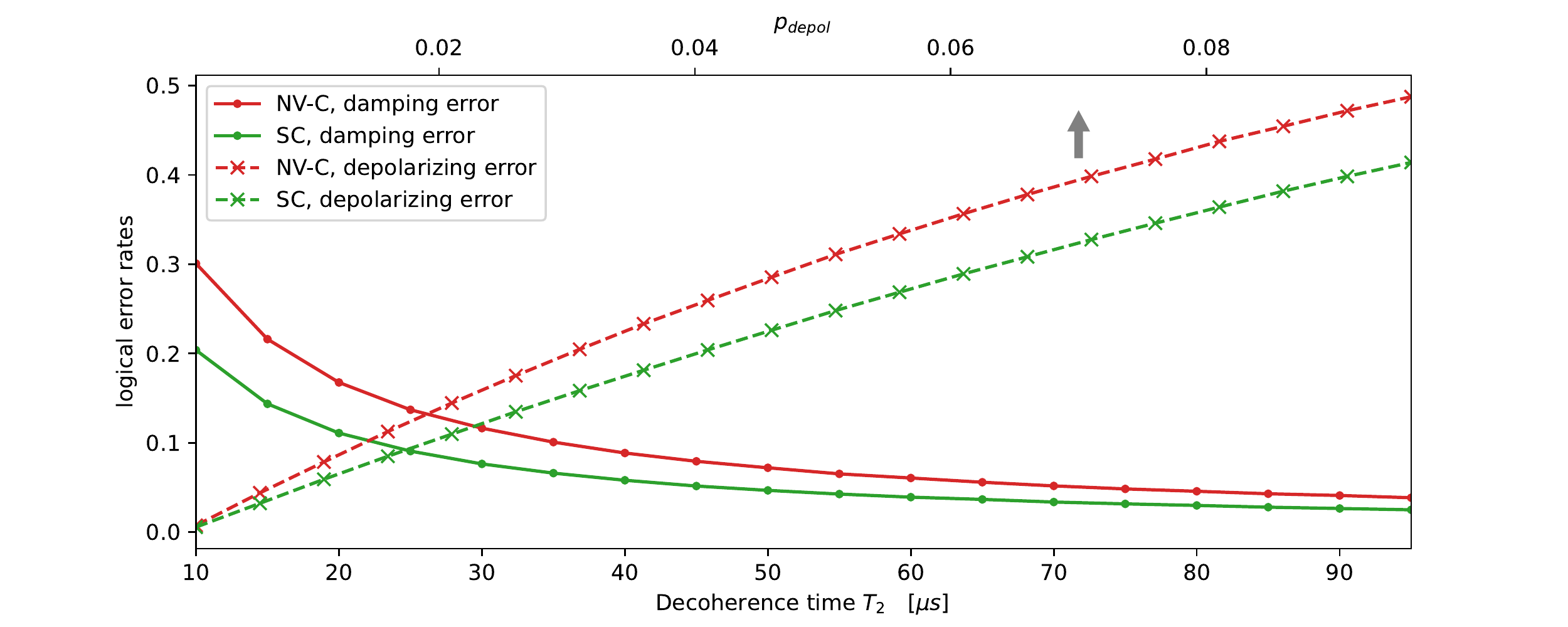}
		\caption{Simulated performance of the three-qubit bit-flip code transpiled for two different hardware platforms, a superconducting processor (SC) and a NV-center register (NV-C). Error rates are averaged over pure input states. Solid lines correspond to the case where we assume perfect gates, only taking errors due to amplitude-phase damping into account ($\mathcal{E}_{\rm apd}(T_1,T_2)$, $\mathcal{E}_{\rm depol}(p_{\rm depol}=0)$, lower axis), while dashed lines represent the case of infinite coherence times and gate errors only, $\mathcal{E}_{\rm apd}(T_1=T_2=\infty)$, $\mathcal{E}_{\rm depol}(p_{\rm depol})$ (upper axis, indicated by the arrow). The implementation on the superconducting processor yields a lower logical error rate, which is due to its lower number of entangling gates. Error bars are smaller than data point markings and thus not displayed.}
		\label{fig:BitFlipcompare}
	\end{figure}
	First, we assume perfect gates and only take errors due to amplitude-phase damping into account, and average over all input states, see figure~\ref{fig:BitFlipcompare} (solid lines). The superconducting processor outperforms the NV-center. This is also the case if only depolarizing gate errors are considered (dashed lines).
	\subsection{Results for the phase-flip code}
\begin{figure}
		\centering
		\includegraphics[width=\textwidth]{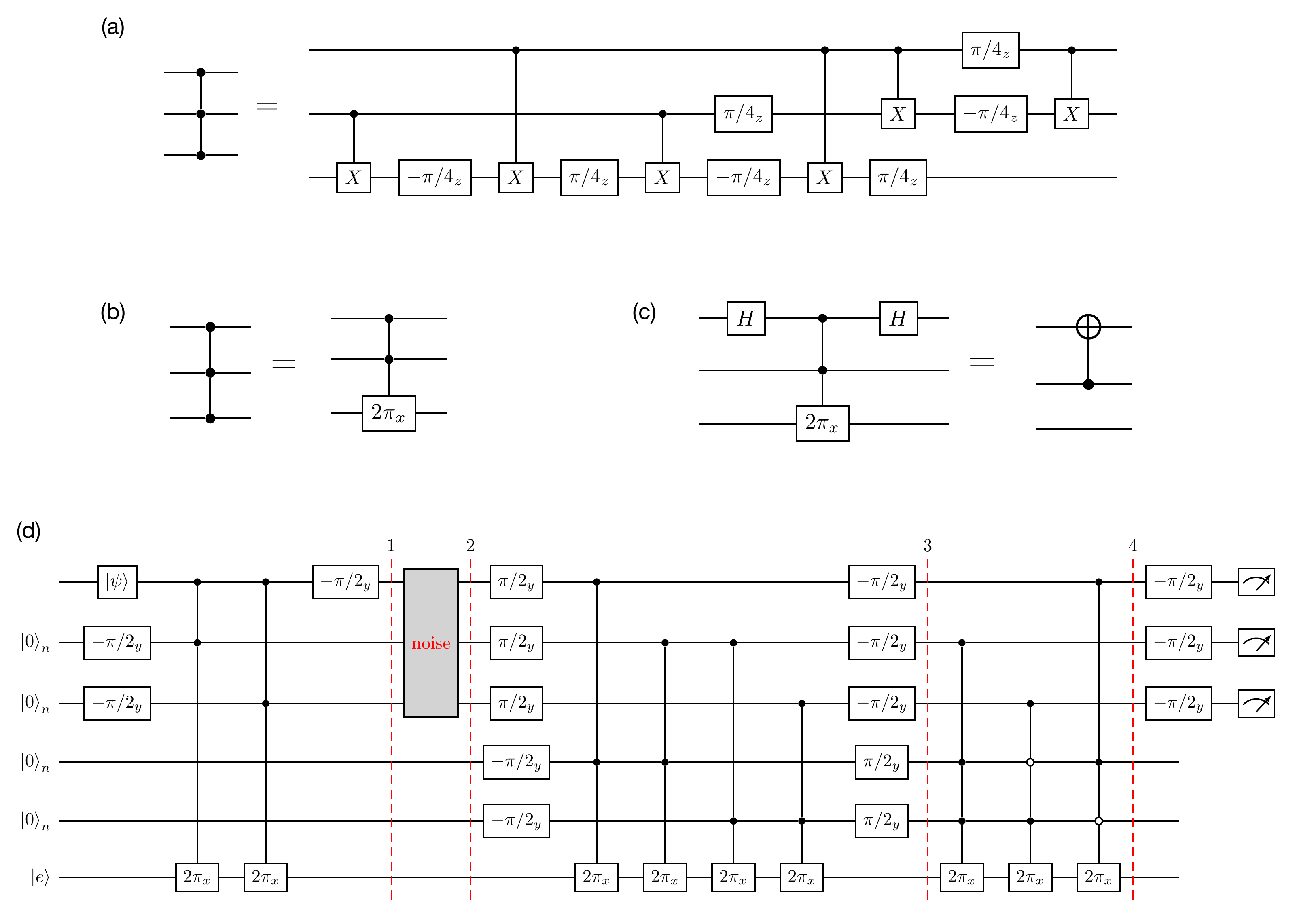}
		\caption{Gate identities for transpiling controlled gates on the superconducting processor and on the NV-center. (a) Transpilation of a CCZ gate on the basis gates provided by superconducting transmon-based processor. For multi-qubit controlled gates, the transpilation to the transmon's nearest-neighbour connectivity leads to an increase in the number of controlled two-qubit gates. The circuit identity holds up to a global phase. (b) The $\textup{C}^n(2\pi_x)$ gate which is native on the electron spin of the NV center is equal to a CCZ gate. (c) A CNOT may be generated from the $\textup{C}^n(2\pi_x)$ gate by adding single qubit rotations. (d) Transpiled phase-flip code given in figure~\ref{fig:repetitioncode}b on a quantum register built by a NV-center in diamond featuring a CSS-like native connectivity displayed in figure~\ref{fig:couplingmaps}b.}
		\label{fig:gate_identites}
	\end{figure}

	Next, we investigate the performance of the phase-flip code. Here, we correct the error unitarily, as depicted in figure~\ref{fig:replace_feedforward}b. 
	In the same manner as described in section~\ref{sec:bitflipperformance}, we use an equal parameter set on both systems, and take the native gates and the native connectivity into account. This allows us to additionally extend the list of basis gates of the NV-center to the uncalibrated $\rm{C}^{(2)}$ and $\rm{C}^{(3)}$ gates which are native on the NV-center register. Their usage has been demonstrated in quantum error-correcting experiments \cite{Waldherr2014}.
	We transpile the circuit onto both platforms. As $\rm{C}^{(2)}$-gates are not native on the superconducting platforms with nearest-neighbour coupling, they have to be translated into $\rm{C}^{(1)}$-gates. With the set of basis gates provided by the transmon qubits, a CCZ-gate transpiles at the cost of 6 $\rm{C}^{(1)}$ gates, see figure~\ref{fig:gate_identites}a \cite{Shende2009,DiVincenzo1998,Barenco1995}. This shows that for multi-qubit controlled gates, the transpilation to the transmon's nearest-neighbour connectivity leads to an increase in the number of controlled two-qubit gates. Also, additional swapping operations are needed due to the limited connectivity of the hexagonal layout, adding up on the number of CNOT gates in the circuit.
	In contrast to this, a CCZ-gate on the electron spin controlled by nuclear spins is native in case of the NV-center. This gate may be used to create arbitrary controlled rotations between the nuclear spins using the gate identities depicted in figure~\ref{fig:gate_identites}b and figure~\ref{fig:gate_identites}c. 

	\begin{figure}
		\centering
		\includegraphics[width=\textwidth]{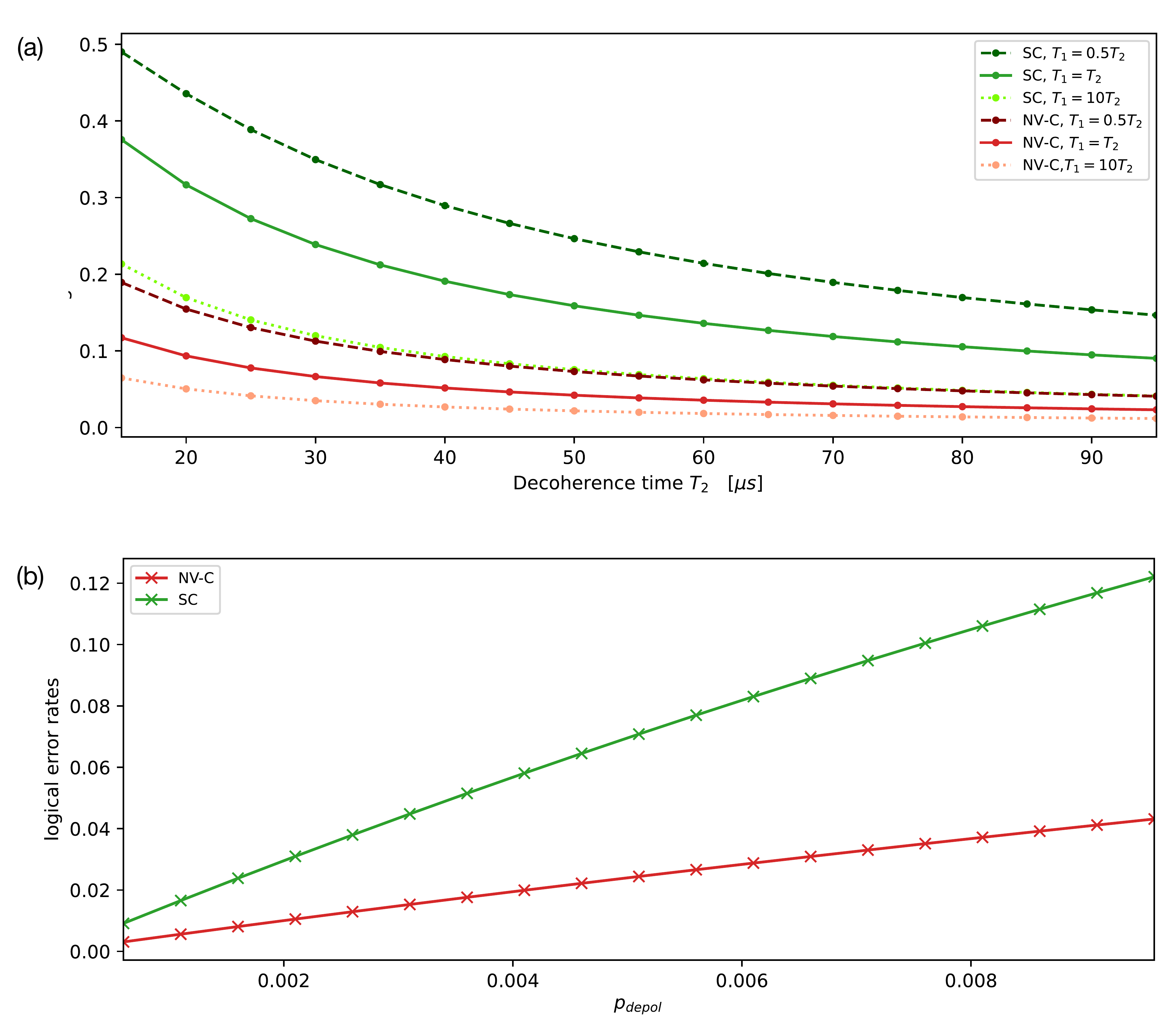}
		\caption{Simulated performance of the three-qubit phase-flip repetition code transpiled for two different hardware platforms, a superconducting processor (SC, shown in green) and a NV-center register (NV-C, shown in red). In (a), we assume perfect gates while only taking errors due to amplitude-phase damping into account ($\mathcal{E}_{\rm apd}(T_2,T_1=\alpha T_2)$, $\mathcal{E}_{\rm depol}(p_{\rm depol}=0)$), plotted over $T_2$ for different ratios  $\alpha=T_1/T_2$. While $\alpha=1$ represents a good approximation for transmon qubits, values for $T_1/T_2$ strongly depend on the NV-center spin type, and we take both $\alpha=0.5$ and $\alpha=10$ into account. Error rates are lower for longer $T_2$ on both platforms, while the NV-center outperforms the SC processor in all cases. In difference to this, we assume infinite coherence times in (b) and take only gate errors into account, $\mathcal{E}_{\rm apd}(T_1=T_2=\infty)$, $\mathcal{E}_{\rm depol}(p_{\rm depol})$. In contrast to the performance of the bit-flip code depicted in figure~\ref{fig:BitFlipcompare}, the NV-center outperforms the superconducting processor in both cases, which is mainly due to the higher number of two-qubit gates in the transpiled code on the superconducting hardware. Note that in both plots, error bars are smaller than data point markings and thus not displayed.}
		\label{fig:phaseflipcompare}
	\end{figure}
	As described in section~\ref{sec:bitflip}, we compare the performance of the two circuits by taking gate errors as well as damping errors into account. Again, we first assume perfect gates and errors due to amplitude-phase damping  ($\mathcal{E}_{\rm apd}(T_2,T_1=\alpha T_2)$, $\mathcal{E}_{\rm depol}(p_{\rm depol}=0)$, see figure~\ref{fig:phaseflipcompare}a, plotted over $T_2$ for different ratios $\alpha=T_1/T_2$. While $\alpha=1$ represents a good approximation for transmon qubits, values for $T_1/T_2$ strongly depend on the NV-center spin type, and we take both $\alpha=0.5$ and $\alpha=10$ into account. In figure~\ref{fig:phaseflipcompare}b, we plot the case of infinite coherence times and imperfect gates. 
	In contrast to the performance of the bit-flip code depicted in figure~\ref{fig:BitFlipcompare}, the NV-center outperforms the superconducting processor in all cases.  A main reason for this difference in performance are the different numbers of multi-qubit gates in the transpiled circuits, which are one of the main bottlenecks for the code performance on both platforms. The transpiled code for the NV-center is depicted in figure~\ref{fig:gate_identites}d. The number of controlled gates is much smaller compared to the transpiled bit-flip code on the transmon processor which transpiles with 35 CNOTs (see figure \ref{fig:transpiledphaseflip_SC} in section \ref{sec:appendix_phaseflipcode}). 

	\section{Conclusions}
	\label{sec:conclusion}
	We compared the performance of quantum error-correcting codes on hardware platforms with different coherence times, connectivity, and native gates: a transmon qubit-based processor and an NV-center quantum register. For this, we used the repetition code correcting either a single bit-flip or a single phase-flip and transpiled them onto both hardware platforms. Additionally, we investigated different methods for replacing the feed-forward operation. Running the code on a superconducting processor, we benchmarked an error model which captures calibrated hardware properties and is thus adaptable to specific quantum processors. We used this model to simulate the impact of amplitude-phase damping and gate errors on the logical error rate for both hardware platforms. While the bit-flip code with post-processing leads to better logical error rates on the superconducting hardware, the phase-flip code with unitary correction shows much better performance on the NV-center register. A predominant reason for this difference in performance lies in the different numbers of multi-qubit gates in the transpiled circuits, which are one of the main bottlenecks for the code performance on both platforms. This strongly indicates that for smaller codes, the quasi-linear layout is advantageous, while for codes involving multi-qubit controlled operations, for instance high-weight parity checks, the native gate set and connectivity of the NV-center allow for a better correction. As multi-qubit controlled operations play an important role in many codes and algorithms, future directions of research could exploit the potential of this property for error correction of CSS-like systems or for designing efficient algorithms tailored to them.
	
	\ack
	We thank Vadim Vorobyov and Thomas Jäger for providing experimental data and for helpful discussions.
	We acknowledge funding from the state of Baden-Württemberg through the Kompetenzzentrum Quantum Computing, project QC4BW.
	
	\appendix
	\section{Kraus operators for amplitude-phase damping}
	\label{sec:appendixkraus}
	
	In this section, we derive a set of Kraus operators for the combined process of amplitude and phase damping $\mathcal{E}_{\rm apd}(\rho)$.
	
	Amplitude damping $\mathcal{E}_{\rm ap}(\rho)$ describes the effect of energy dissipation into the environment of a qubit \cite{Chirolli2008,Krantz2019}. Its operators couple to the x-y plane of the Bloch sphere and cause a longitudinal relaxation to a certain steady state with 
	\begin{equation}
		E_{\rm ad,0}=\left( \begin{array}{rr} 1 & 0\\ 0 & \sqrt{1-p_{\rm ad}} \end{array}\right), \quad
		E_{\rm  ad,1}=\left( \begin{array}{rr} 0 & \sqrt{p_{\rm ad}}\\ 0 & 0	\end{array}\right),
	\end{equation}
	where $p_{\rm ad}$ represents the probability for the qubit to loose an excitation into the environment.
	
	Phase damping $\mathcal{E}_{\rm pd}(\rho)$ describes the loss of information about the relative phases between the energy eigenstates into an environment, but does not affect the eigenstates themselves. In a simple model, it may be described by random phase kicks with a Gaussian-distributed variable. One possible set of its Kraus operators is given by  
	\begin{equation}
		E_{\rm pd,0}=\left( \begin{array}{rr}
			1 & 0\\
			0 & \sqrt{1-p_{\rm pd}}
		\end{array}\right), \quad
		E_{\rm pd,1}=\left( \begin{array}{rr}
			0 & 0\\
			0 & \sqrt{p_{\rm ad}}
		\end{array}\right).
	\end{equation}
	
	The set of Kraus operators for two combined quantum operations $\mathcal{E}_{\rm A}$ and $\mathcal{E}_{\rm B} (\rho)$ acting non-trivially on the same Hilbert space $\mathcal{H}_{\rm AB}$ and described by a set of Kraus operators $\left\{ E_{\rm A,\textit{j}}\right\}$, $\left\{E_{\rm B,\textit{i}}\right\}$ may be obtained according to
	\begin{equation}
	    \mathcal{E}_{\rm A} \circ \mathcal{E}_{\rm B} (\rho)=\sum_j E_{\rm A,\textit{j}}\left(\sum_i E_{\rm B, \textit{i}}\rho E^\dagger_{\rm B, \textit{i}}\right) E^\dagger_{\rm A,\textit{j}}
	\end{equation}
	Calculating $\mathcal{E}_{\rm apd}(\rho)=\mathcal{E}_{\rm pd} \circ \mathcal{E}_{\rm ad} (\rho)$ leads to the set of Kraus operators (also given in \cite{Ghosh2012})
	\begin{eqnarray}
		E_{\rm apd,0}&=\left( \begin{array}{rr}
			1 & 0\\
			0 & \sqrt{1-p_{\rm ad}}\sqrt{1-p_{\rm pd}}
		\end{array}\right), 
		E_{\rm apd,1}=\left( \begin{array}{rr}
			0 & \sqrt{p_{\rm ad}}\\
			0 & 0
		\end{array}\right),\\
		E_{\rm apd,2}&=\left( \begin{array}{rr}
			0 & 0\\
			0 & \sqrt{1-p_{\rm ad}}\sqrt{p_{\rm pd}}
		\end{array}\right)
	\end{eqnarray}
	where we may relate $p_{\rm ad}, p_{\rm pd}$ to the relaxation time $T_1$, the dephasing time $T_2$ and the gate time $\Delta t$ with 
	\begin{equation}
		\sqrt{1-p_{\rm ad}}= e^{\Delta t/2T_1}, \sqrt{1-p_{\rm ad}}\sqrt{1-p_{\rm pd}}= e^{\Delta t/2T_2}.
	\end{equation}
	While the reverse order $\mathcal{E}'_{\rm apd}(\rho)=\mathcal{E}_{\rm ad} \circ \mathcal{E}_{\rm pd} (\rho)$ leads to a different set of Kraus operators, their effect on the quantum state is identical, thus $\mathcal{E}_{\rm apd}(\rho)= \mathcal{E}'_{\rm apd}(\rho)$. This can be seen by calculating their unique Choi representation $J(\rho)$, which is related to the Kraus representation with 
	\begin{equation}
	    J(\rho)= \sum_k \rm{vec}(E_k)\rm{vec}(E^\dagger_k)
	\end{equation}
	where $\rm vec(\cdot)$ represents the vectorization operation \cite{Watrous2018}. With this, we calculate
	\begin{eqnarray}
	 J(\mathcal{E}_{\rm apd}(\rho))&=
	 J(\mathcal{E}'_{\rm apd}(\rho))=\\
	 &\left(\begin{array}{cccc}
	      1 & 0 & 0 & \sqrt{1-p_{\rm ad}}\sqrt{p_{\rm pd}} \\
	      0 & p_{\rm ad} & 0 & 0 \\
	      0 & 0 & 0 & 0 \\
	      \sqrt{1-p_{\rm ad}}\sqrt{p_{\rm pd}} & 0 & 0 & 1-p_{\rm ad}
	 \end{array}\right).
	\end{eqnarray}

    \clearpage
	\section{Transpiled phase-flip code on the superconducting hardware}
	\label{sec:appendix_phaseflipcode}
	\begin{figure}[!h]
		\centering
		\includegraphics[width=\textwidth]{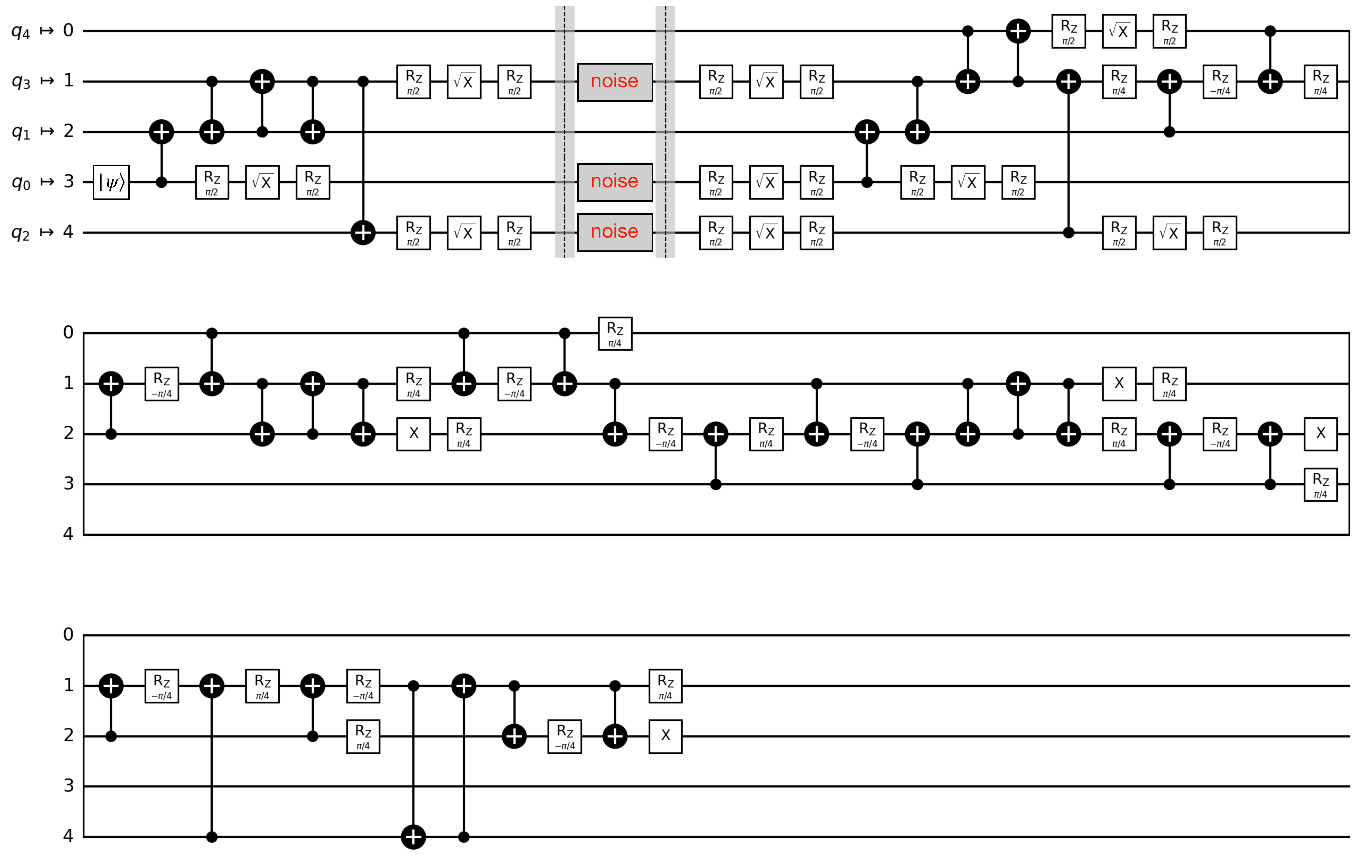}
		\caption{Transpiled phase-flip code on the superconducting processor. The connectivity is given by a bidirectional graph with the nodes $(0,1,2,3,4)$ and the vertices $(0, 1), (1, 0), (1, 2), (2, 1), (2, 3), (3, 2),(1,4),(4,1)$. The initial layout is indicated by the mapping of the physical qubits $\left\{q_i \right\}$ to the nodes.}
		\label{fig:transpiledphaseflip_SC}
	\end{figure}

	\clearpage
	\section*{References}
	\bibliography{bibliography.bib}
	
\end{document}